\documentclass[epj]{svjour}
\usepackage{times}
\usepackage{graphicx}
\usepackage{dcolumn}
\usepackage{bm}
\usepackage[colorlinks,breaklinks,citecolor=blue,linkcolor=blue,urlcolor=blue]{hyperref}
\usepackage[T1]{fontenc} 
\usepackage{amsfonts}
\usepackage{amsmath}

\begin{document}

\title{Periodic driving induced helical Floquet channels with ultracold atoms in momentum space}
\author{Teng Xiao\inst{1}, Dizhou Xie\inst{1}, Wei Gou\inst{1}, Tao Chen\inst{1}, Tian-Shu Deng\inst{2}, Wei Yi\inst{3,4}, Bo Yan\inst{1,\dagger} 
}                     
%
%
\institute{Interdisciplinary Center of Quantum Information, State Key Laboratory of Modern Optical Instrumentation, and Zhejiang Province Key Laboratory of Quantum Technology and Device of Physics Department, Zhejiang University, Hangzhou 310027, China \and Institute for Advanced Study, Tsinghua University, Beijing, 100084, China \and CAS Key Laboratory of Quantum Information, University of Science and Technology of China, Hefei 230026, China \and CAS Center For Excellence in Quantum Information and Quantum Physics, Hefei 230026, China \and $^\dagger$ yanbohang@zju.edu.cn }
\date{Received: date / Revised version: date}
%
\abstract{
Employing the external degrees of freedom of atoms as synthetic dimensions renders easy and new accesses to quantum engineering and quantum simulation. As a recent development, ultracold atoms suffering from two-photon Bragg transitions can be diffracted into a series of discrete momentum states to form a momentum lattice. Here we provide a detailed analysis on such a system, and, as a concrete example, report the observation of robust helical Floquet channels, by introducing periodic driving sequences. The robustness of these channels against perturbations is confirmed, as a test for their topological origin captured by Floquet winding numbers. The periodic switching demonstrated here serves as a testbed for more complicated Floquet engieering schemes, and offers exciting opportunities to study novel topological physics in a many-body setting with tunable interactions.
} 
\authorrunning{ }
\titlerunning{ }
\maketitle

\section{Introduction}

First proposed by Feynman, quantum simulation offers an intriguing prospect toward an efficient understanding of complicated quantum many-body systems and difficult physical models in a quantum-mechanical fashion~\cite{Lewenstein2007,Jaksch2007,Georgescu2014}. A common practice in quantum simulation is the so-called bottom-up approach, i.e., mapping the already known Hamiltonian to a carefully designed simulator, which provides opportunities for gaining insights of complex phenomena in diverse contexts, ranging from condensed-matter physics to quantum chemistry and nuclear physics. Concrete examples include quantum phase transitions in Hubbard models~\cite{Greiner2002,Jordens2008,Schneider2008,Baier2016}, high-$T_c$ superconductivity~\cite{Hofstetter2002,Lee2006}, quantum magnetism and chaos~\cite{Struck2011,Greif2013,Russell2015,Luca2016}, topological order~\cite{Micheli2006,Aguado2008,You2010}, and lattice gauge theories~ \cite{Buchler2005,Byrnes2006,Glaetzle2014,Mezzacapo2015,Luca2019}. Benefiting from the rapid development of experimental techniques of quantum manipulation, quantum simulation has become quite feasible, and has attracted much interest in a wide range of platforms including ultracold atoms~\cite{Bloch2012}, trapped ions~\cite{Blatt2012}, photons~\cite{Guzik2012} and superconducting circuits~\cite{Houck2012}.

With highly controllable experimental parameters, ultracold atoms in optical lattices provide an excellent toolbox for studying exotic physics in condensed-matter systems. During the last decade, one has witnessed revolutionary advances such as the preparation of artificial solids~\cite{Bloch2012,Gross2017}, the observation of new phases of matter~\cite{Henkel2010,Julian2017,Leonard2017}, the simulation of many-body systems~\cite{Baumann2010,Schreiber2015,Choi2016,Smith2016} and topological matter~\cite{Aidelsburger2013,Miyake2013,Jotzu2014,Duca2015,Stuhl2015,Flaschner2016,Cooper2019}. In these studies, optical lattices play a central role in simulating electrons in solids using neutral atoms. However, the traditional optical lattices are generally implemented in real space, where the small spacing between two adjacent sites ($\sim \mu \rm m$) makes local control on the single-site level rather complicated~\cite{Muller2018}. To overcome the difficulty, the idea of synthetic dimension is proposed~\cite{Boada2012,Celi2014,Ozawa2019} by using the internal or external degrees of freedom of an atom to form a lattice structure~\cite{Zeng2015,Livi2016,Luca2017,Wall2015,Cai2019}. Such a concept can not only extend the system to higher dimensions~\cite{Price2015,Zilberberg2018},  but also provides outstanding control capabilities for quantum engineering and quantum simulation.

In this work, we implement the synthetic dimension with external momentum states of ultracold atoms. Our scheme also enables us to demonstrate helical Floquet channels in the one-dimensional momentum lattice~\cite{Budich2017}, where atoms on different sublattice sites are locked into a leftward (rightward) unidirectional motion along the lattice. While the helical Floquet channels derive from the winding of Floquet quasienergy bands, we explicitly demonstrate their robustness against perturbations, as a test to their topological origin.

The paper is organized as follows. In Sec.~\ref{sec2}, we show details on the implementation of the synthetic momentum lattice. In Sec.~\ref{sec3}, we show that by precisely modulating the hopping amplitudes between each pair of lattice sites, we effectively implement a discrete-time quantum-walk dynamics. Then, in Sec.~\ref{sec4}, we demonstrate the robustness of helical Floquet channels to noises. We conclude in Sec.~\ref{sec5}.

\section{Synthetic dimension with momentum states}\label{sec2}

The synthetic dimension in our experiment is formed by a set of discrete momentum states of ultracold $^{87}$Rb atoms. The scheme for momentum lattice is first proposed in Ref.~\cite{Gadway2015}, and subsequently experimentally realized~\cite{Meier2016,Xie2019}. We briefly summarize the scheme as follows. We trap $\sim 6\times 10^4$ $^{87}$Rb atoms (condensed at the momentum $p= 0$) in an optical dipole trap with trapping frequencies $\sim 2\pi\times(120, 40, 120)~\rm Hz$. The atoms are subject to a series of Bragg lasers (at the wavelength $\lambda=1064$nm), and are coupled to higher momentum states. Due to the momentum conservation, the allowed momenta are discretely distributed as $p = 2np_r$ ($n\in\mathbb{Z}$), where $p_r$ is the photon-recoil momentum of the Bragg lasers. If we label the momentum state $p=2np_r$ with $|n\rangle$, a synthetic one-dimensional (1D) lattice chain is constructed, whose length is determined by the number of coupled momentum states.

\subsection{Effective tight-binding Hamiltonian}

\begin{figure}[]
\centering
\includegraphics[width=0.44\textwidth]{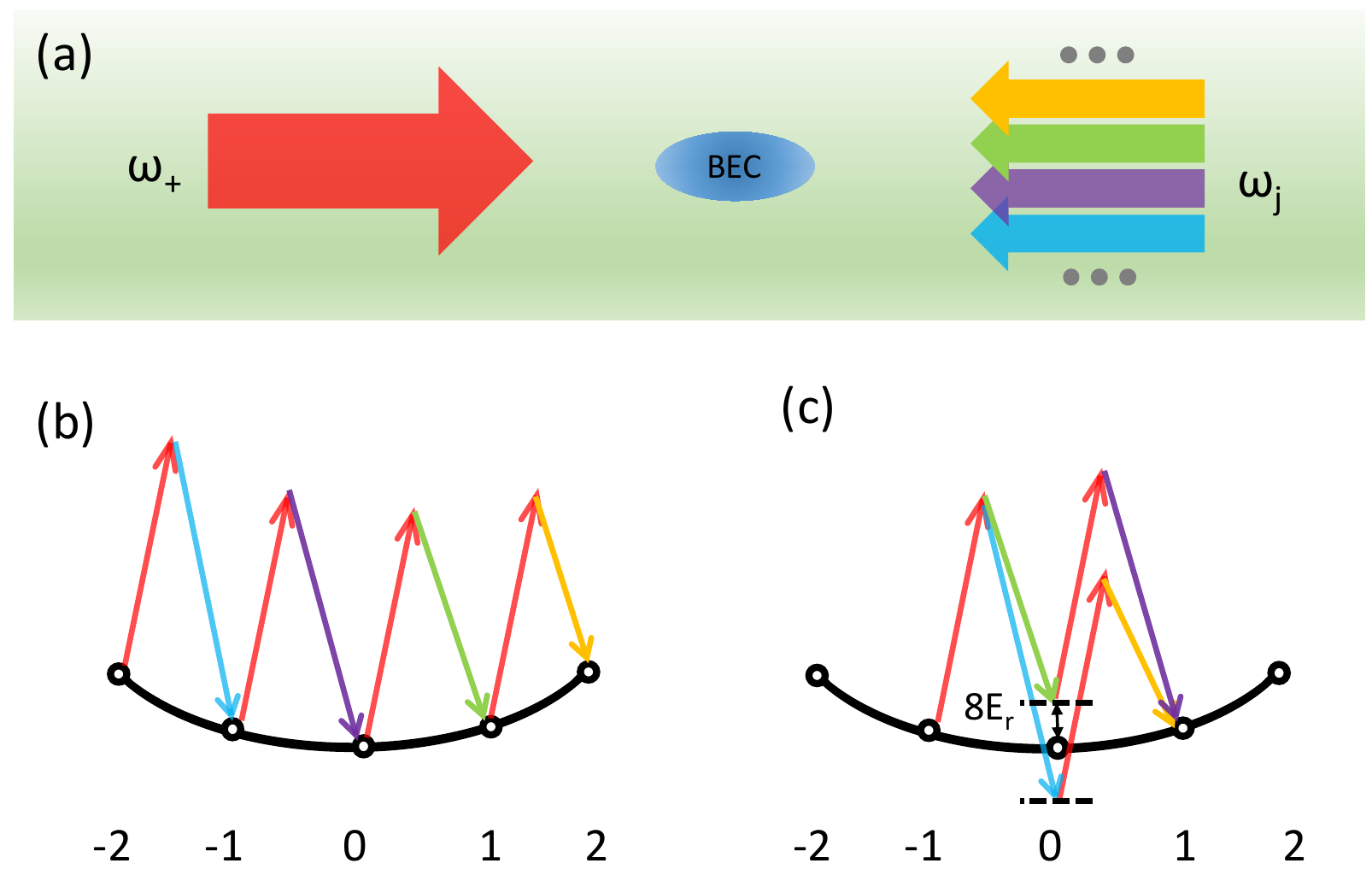}
\caption{\label{fig1} Schematic of the experimental setup.
(a) A $^{87}$Rb Bose-Einstein condensate interacts with two counterpropagating Bragg laser beams. The right-propagating beam has a single frequency $\omega_+$, while the frequencies $\omega_j$ of left-propagating beam are precisely programmed with acoustic optical modulators to address different momentum states.
(b) Resonant couplings triggered by two-photon Bragg transitions. The laser pair $\{\omega_+, \omega_j\}$ resonantly couples the momentum states $|j\rangle$ and $|j+1\rangle$.
(c) One example of four-photon transitions for $n=0$ and $\ell=1$. Now the coupling $|-1\rangle\leftrightarrow|+1\rangle$ can be induced either by laser pairs with frequencies $\{\omega_+, \omega_{-1}; \omega_+, \omega_{0}\}$ or by those with $\{\omega_+, \omega_{1}; \omega_+, \omega_{-2}\}$. The two four-photon processes have the two-photon detunings of $\pm 8E_r$, respectively.
}
\end{figure}

The Bragg lasers in our experiment are applied as schematically illustrated in Fig.~\ref{fig1}(a). One right-propagating laser beam has a single frequency $\omega_+$, and the frequencies of the left-propagating beam are chosen as $\omega_j = \omega_+-4(2j+1)E_r/\hbar$ ($j\in\mathbb{Z}$), where $E_r = p_r^2/2\mu$ is the photon-recoil energy ($\mu$ is the atomic mass). In our case, $E_r=h\times 2.03$kHz. Each pair of laser beams $\{\omega_+,\omega_j\}$ triggers a resonant two-photon Bragg transition to couple the momentum states $|j\rangle\leftrightarrow|j+1\rangle$. Following the treatment in~\cite{Gadway2015,Giese2013}, the full Hamiltonian under the interaction picture reads \cite{Gou2020}
\begin{equation}\label{eq:full}
\hat{H}_{\rm full} =\sum\limits_n \sum_{j\in \mathbb{Z}} \frac{\hbar\tilde\Omega_j}{2}e^{i8(n-j)E_rt/\hbar}|n+1\rangle\langle n|  + \rm{H.c.},
\end{equation}
where $\tilde\Omega_j$ is the effective Rabi frequency of the two-photon transition $|j\rangle\leftrightarrow|j+1\rangle$ from the laser pair $\{\omega_+,\omega_j\}$. $\tilde\Omega_j$ can be easily adjusted by tuning the intensity of the relevant laser beam.

Re-organizing Eq.~(\ref{eq:full}) in terms of different off-resonant detunings, we have
\begin{equation}
\hat{H}_{\rm full} = \sum_{\ell}  H^{[\pm2\ell]} e^{\pm i8\ell E_rt/\hbar},
\end{equation}
with $\ell \in \mathbb{N}$, and
\begin{equation}
H^{[+2\ell]} = \sum_n \frac{\hbar\tilde\Omega_{n-\ell}}{2} |n+1\rangle \langle n| + \sum_n\frac{\hbar\tilde\Omega_{n+\ell}}{2} |n\rangle \langle n+1|,
\end{equation}
where $H^{[-2\ell]} = H^{[+2\ell]\dagger}$.
Apparently, all the resonant two-photon Bragg diffractions are described by the term
\begin{equation}\label{eq:4}
H^{[0]} = \sum_n \frac{\hbar\tilde\Omega_n}{2} |n+1\rangle\langle n| + \rm{H.c.}.
\end{equation}
Figure~\ref{fig1}(b) shows the coupling schemes corresponding to the approximation Hamiltonian $H^{[0]}$. The $2\ell$-th order terms are contributed by the $\{\omega_+, \omega_{n-\ell}\}$ and $\{\omega_+,\omega_{n+\ell}\}$ laser pairs, which induce the transition $|n\rangle\leftrightarrow|n+1\rangle$ with a detuning of $\pm 8\ell E_r$, respectively. These off-resonant terms give the higher-order effect, and sometimes can be neglected.

To clearly see the effect of the higher-order  off-resonant terms, we approximate the above time-dependent full Hamiltonian, up to the second-order corrections, by an equivalent time-independent one~ \cite{Giese2013,Goldman2014}
\begin{equation}\label{eq6}
\hat{H} = H^{[0]} + \sum\limits_{\ell\in N^+} \frac{[H^{[+2\ell]}, H^{[-2\ell]}]}{4(2\ell)E_r} .
\end{equation}
By letting $t_j = \hbar\tilde\Omega_j/2$, we have
\begin{equation}
\begin{split}
\hat{H} = &\sum_n (t_n^{(1)}|n+1\rangle\langle n| + {\rm h.c.}) + \sum_n (t_n^{(2)} |n+1\rangle\langle n-1|+{\rm h.c.})  \\
&+ \sum_n\Delta_n |n\rangle\langle n|.
\end{split}
\end{equation}
with $t_n^{(1)} = t_n$, and
\begin{eqnarray}
t_n^{(2)} &=& \sum\limits_{\ell\in \mathbb{N}^{+}}\frac{1}{8\ell E_r}(t_{n-\ell}t_{n+\ell-1} - t_{n+\ell}t_{n-\ell-1}) , \label{eq11}\\[0.5em]
\Delta_n & = &  \sum\limits_{\ell\in \mathbb{N}^{+}}\frac{1}{8\ell E_r}(t^2_{n+\ell}+t^2_{n-\ell-1}-t^2_{n-\ell}-t^2_{n+\ell-1}).
\end{eqnarray}
Here the $t_n^{(2)}$ and $\Delta_n$ terms results from the four-photon processes. The $\ell$-th term includes two four-photon processes: one is with the laser pair of $\{\omega_+, \omega_{n+\ell-1}; \omega_+, \omega_{n-\ell}\}$, the another is with the laser pair of   $\{\omega_+, \omega_{n+\ell}; \omega_+, \omega_{n-\ell-1}\}$. They both resonantly couple $|n-1\rangle\leftrightarrow|n+1\rangle$ via the four-photon processes but with two-photon detunings of $\pm 8\ell E_r/\hbar$, respectively. One example is illustrated in Figure~\ref{fig1}(c)  for $n=0$ and $\ell=1$. Because the different signs of the detuning for the two processes, they actually cancel each other for $t_n^{(2)}$ and $\Delta_n$ in some sense. These make such effects even less important. If we neglect the terms with denominators larger than $8E_r$ ($\sim h\times16.2~\rm{kHz}$ in our experiment), we arrive at an effective tight-binding Hamiltonian
\begin{equation}\label{eq:eff}
\hat{H}_{\rm eff} = \sum_n t_n (|n+1\rangle\langle n| + |n\rangle\langle n+1|).
\end{equation}

\subsection{Role of the off-resonant, time-dependent couplings}

\begin{figure}[]
\centering
\includegraphics[width= 0.4\textwidth]{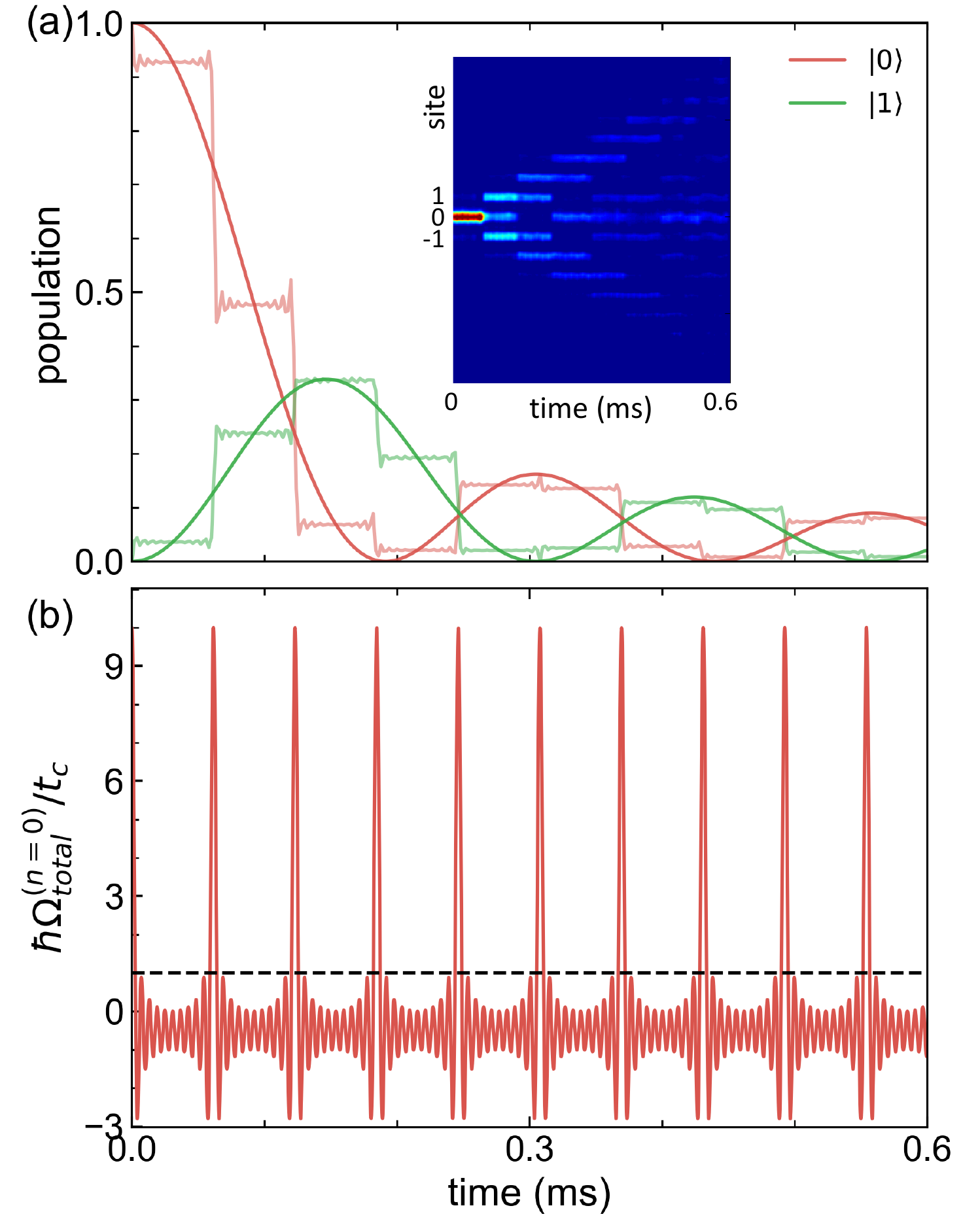}
\caption{\label{fig2}(Color online) Role of the time-dependent terms in the full Hamiltonian.
(a) Time evolution of populations in sites $|0\rangle$ (red) and $|1\rangle$ (green). The simulations are performed for a 21-site momentum-state lattice. The results from the full Hamiltonian are plotted in light colors, while lines in deep colors are for the effective Hamiltonian.
The inset shows the experimentally measured time evolution of the population in each site with similar parameters.
(b) The time-dependent coupling strength $\Omega_{\rm total}^{(n=0)}$ for transition $|0\rangle\leftrightarrow|1\rangle$. The black dashed line indicates the constant coupling strength $t_c$ in the effective model. In simulations here, we choose a uniform coupling strength, i.e., $t_j = t_c = h\times 1.0 ~{\rm kHz}$ ($j\in\mathbb{Z}$), and the atoms are initially prepared in site $|0\rangle$.}
\end{figure}

In the above derivation of the effective Hamiltonian (\ref{eq:eff}), we directly apply the perturbation expansion (\ref{eq6}) for a Floquet quantum system~\cite{Goldman2014} to average out the time-dependent terms. To have a deeper insight on the impact of the time-dependent terms, we compare results from numerical simulations with the full Hamiltonian (\ref{eq:full}) with those from the effective one (\ref{eq:eff}). Figure~\ref{fig2}(a) shows the numerical time evolution of the populations in site $|0\rangle$ and site $|1\rangle$ on a 21-site momentum state lattice when $t_j$ are constant $t_j=t_c=h\times$1kHz. The effect from the time-dependent terms is clearly visible. A series of step-like jumps in the atom population occurs in the full-Hamiltonian simulation. The profiles of the evolutions fit well with the smooth lines obtained with the effective Hamiltonian.

The step-like jump phenomenon originates from the interference of the higher-order off-resonant couplings which are neglected in effective Hamiltonian (\ref{eq:eff}). This is easily understood from the expression of the time-dependent coupling strength for transition $|n\rangle\leftrightarrow|n+1\rangle$ is
\begin{equation}
\hbar\Omega_{\rm total}^{(n)}=\sum_{j\in \mathbb{Z}} t_j{\rm exp}{[i8(n-j)E_rt/\hbar]}.
\end{equation}
Figure~\ref{fig2}(b) gives an example for the case $n=0$. The interference between different frequency components results in a series of periodic peaks. From Fig.~\ref{fig2}(a), one can find that, the hopping between two sites occur mostly at times when these sharp peaks appear. The time interval between two adjacent peaks is $\tau_{\rm p}\sim \hbar/8E_r$. For a smaller hopping strength $t_c$, more steps with smaller gaps in between are involved during one period $T = \hbar/t_c$, making the time evolution more smooth which fit better with simulation results from the effective Hamiltonian. However, when $\tau_{\rm p} \sim T$, the number of steps in each period is small, and the effective Hamiltonian is not a good approximation any more. This is consistent with the condition $t_c \ll 8E_r$, under which the time-independent Hamiltonian~\cite{Goldman2014} for a Floquet system is derived. The insert of Fig.~\ref{fig2} (a) shows one experimental data with the similar parameters, the step-like behavior is clearly observed.

\section{Quantum walk with periodic driving}\label{sec3}

\begin{figure}[]
\includegraphics[width= 0.45\textwidth]{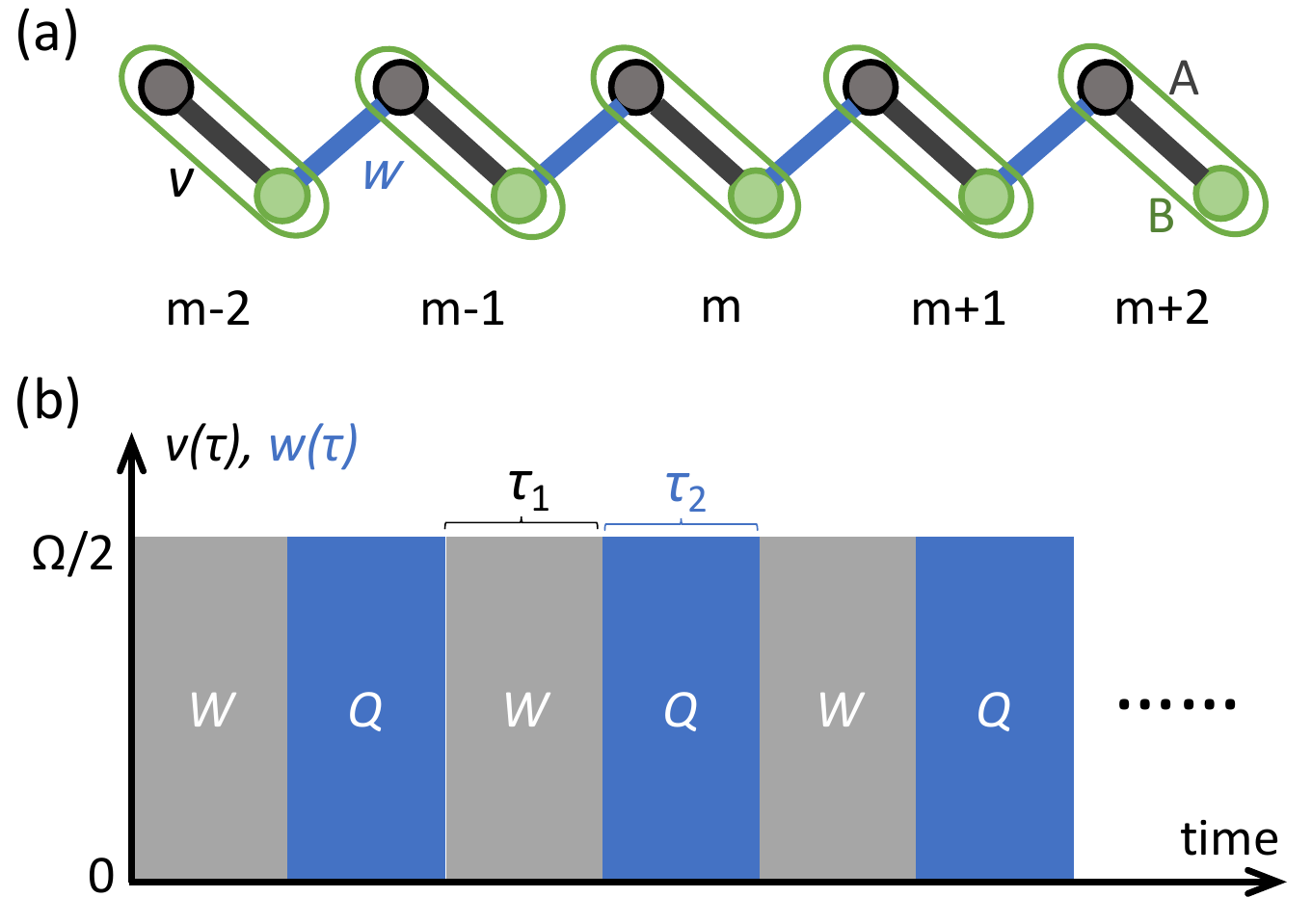}
\centering
\caption{\label{fig3}(Color online) Realization of a discrete-time quantum walk in the momentum state lattice.
(a) Mapping of the momentum state lattice to an SSH model. The inter- and intra-cell hopping amplitudes are set as $w$ and $v$ respectively.
(b) Implementation of a quantum walk by periodically modulating the hopping amplitudes. The inter- and intra-cell hoppings are alternatively switched on and off. The corresponding evolution operators are labelled by $Q$ and $W$ respectively, and the durations for each operation are $\tau_1$ and $\tau_2$.}
\end{figure}

An important advantage of the synthetic dimension with the momentum states is the highly tunable local parameters. In the effective Hamiltonian, the tunnelings, $t_n$, can be independently varied, enabling us to engineer some interesting models in condensed-matter physics. For example, by tuning the odd and even hopping amplitudes to be different, i.e., $v$ and $w$ in Fig.~\ref{fig3}(a), the momentum-state chain can be mapped into the Su-Schrieffer-Heeger (SSH) model~\cite{Meier2016a} as well as its variants~\cite{Xie2019}. To implement the quantum-walk scheme, the odd and even hoppings are alternatively switched on and off, as shown in Fig.~\ref{fig3}(b). Both the intra- and inter-cell hoppings have an amplitude of $\hbar\Omega/2$, and the time durations are $\tau_1$ and $\tau_2$, respectively.  Now the periodically driven Hamiltonian for the system in Fig.~\ref{fig3}(a) can be written as \cite{Xie2019b}
\begin{figure*}[tbp]
\centering
\includegraphics[width= 0.85\textwidth]{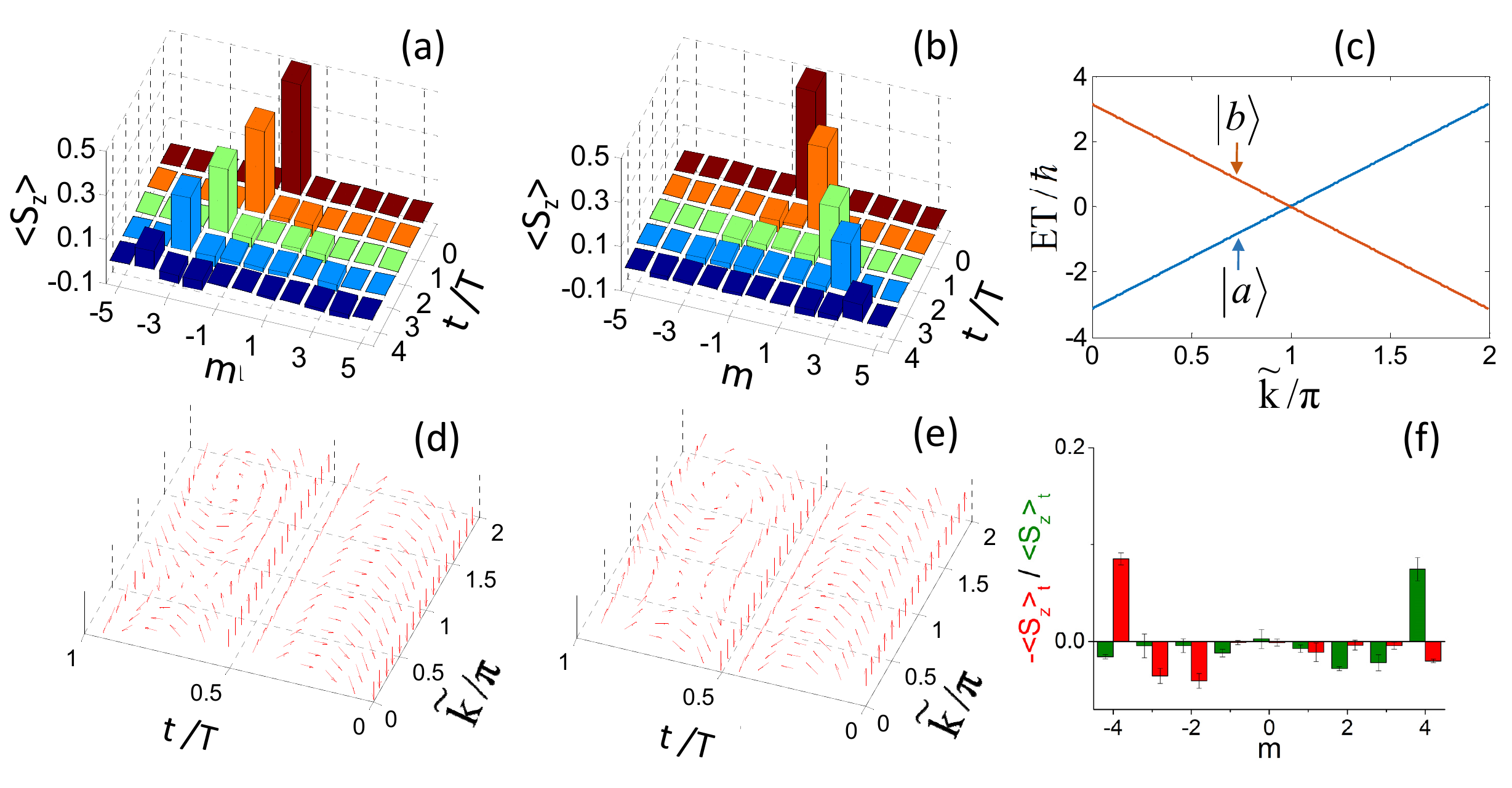}
 \caption{(Color online) Observation of helical Floquet channels with $(\theta_1,\theta_2)=(\pi/2,\pi/2)$. (a) Measured $-\langle S_z\rangle_t$ shows a unidirectional leftward flow for atoms initialized in $|0,b\rangle$. (b) Measured $\langle S_z\rangle_t$ shows a unidirectional rightward flow for atoms initialized in $|0,a\rangle$. (c) Quasienergy dispersion for $(\theta_1,\theta_2)=(\pi/2,\pi/2)$. (d)(e) Space-time spin textures corresponding to dynamics in (a)(b), respectively. (f) Measured $\pm\langle S_z\rangle_t$ for $N_{\rm step}=4$. Red (black) bars correspond to $-\langle S_z\rangle_t$ ($\langle S_z\rangle_t$ ) for atoms initialized in $|0,b\rangle$ ($|0,a\rangle$).
}\label{fig:fig4}
\end{figure*}
\begin{equation}
\begin{split}
{\hat H}=&v(t)\sum\limits_{m=1}^{N}(\left| m,B \right\rangle \left\langle m,A \right| + {\rm h.c.})  + \\
&+w(t)\sum\limits_{m=1}^{N-1}(\left| m+1,A \right\rangle \left\langle m,B \right| + {\rm h.c.}).
\end{split}
\end{equation}
This formula is similar with the Hamiltonian of the SSH model but with time-dependent hoppings. When only turning on the intra-cell hoppings, i.e., $v=\hbar\Omega/2$ and $w=0$, the time evolution of the system is determined by the $W$ operator which induces intra-cell dynamics as
\begin{eqnarray}
W\left| m,A \right\rangle &=&\cos\theta_1\left| m,A \right\rangle+i \sin\theta_1\left| m,B \right\rangle,\\
W\left| m,B \right\rangle &=&\cos\theta_1\left| m,B \right\rangle+i \sin\theta_1\left| m,A \right\rangle,
\end{eqnarray}
with $\theta_1=\Omega \tau_1/2$. Similarly, the $Q$ operator leads to
\begin{eqnarray}
Q\left| m,A \right\rangle &=& \cos\theta_2\left| m,A \right\rangle+i\sin\theta_2\left| m-1,B \right\rangle,\\
Q\left| m,B \right\rangle &=& \cos\theta_2\left| m,B \right\rangle+i\sin\theta_2\left| m+1,A \right\rangle,
\end{eqnarray}
where $\theta_2=\Omega \tau_2/2$. Once we choose the duration $\tau_1$ such that $\theta_1=\pi/4$, such a scheme forms a standard discrete-time quantum walk. The step of $W$ operation creates the intra-cell superpositions while the cell-dependent $Q$ step shifts these states further.

\section{Robust helical Floquet channels}\label{sec4}
\begin{figure*}[]
\includegraphics[width= 0.9\textwidth]{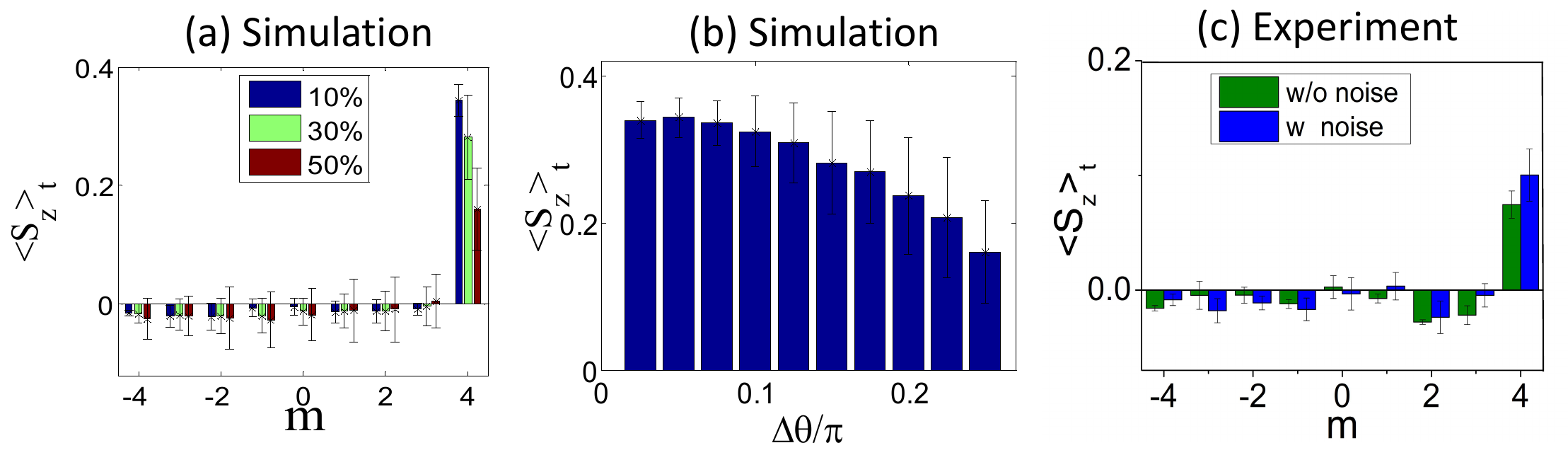}
\centering
\caption{(Color online) Robustness of the helical Floquet channels against random noise. (a) and (b) show the numerical simulations of the helical Floquet channel in the presence of increasing noise.
For numerical calculations, we generate $100$ sets of parameters, where $(\theta_1,\theta_2)$ in each time step are randomly picked in the interval $(\pi/2-\Delta\theta,\pi/2+\Delta\theta)$. We then let the initial state evolve under the full Hamiltonian for the duration of $N_{\rm step}=4$ steps. In (a), we plot the averaged $\langle S_z\rangle_t$ of the last step, for $\Delta\theta=\pi/2\times10\%$ (blue), $\Delta\theta=\pi/2\times30\%$ (green), and $\Delta\theta=\pi/2\times50\%$ (brown), respectively. In (b), we show the averaged peak in $\langle S_z\rangle_t$ of the last step, which decreases with increasing noise level. (c) shows the experimental data with and without noise. The data is averaged for 10 times with 10\% random noise.
}\label{fig:fig5}
\end{figure*}

The high tunability of our configuration enables us to observe robust helical Floquet channels in the 1D momentum lattice~\cite{Budich2017,Mukherjee2018}. To observe the phenomenon, we evolve the system under the Floquet operator $U_h=Q(\pi/2)W(\pi/2)$, implemented with $T\approx 0.44$ms ($\Omega=2\pi\times 2.3(1)$kHz $\ll 8E_r/\hbar$) and $t_w=t_q$. We then measure $\langle S_z\rangle_t$ as a function of $m$ at each time step for different initial states. Here we define the pseudo-spin operators $S_\beta$ ($\beta=x,y,z$) with $S_\beta=\frac{1}{2}|m,\mu\rangle\sigma^{\mu\nu}_\beta\langle m,\nu|$, where $\mu,\nu\in\{a,b\}$ and $\sigma^{\mu\nu}_\beta$ is the corresponding Pauli matrix element~\cite{Budich2017}.

As shown in Fig.~\ref{fig:fig4}(a)(b), atoms initialized in $|m=0,b\rangle$ ($|m=0,a\rangle$) propagate to the left (right) in the quantum-walk dynamics. The unidirectional flow is clearly shown in the measured $\langle S_z\rangle_{t}$. On the microscopic level, such a unidirectional atom flow is generated by a series of alternating spin-flipping processes and spin-dependent shifts over half the lattice spacing~\cite{Budich2017}. Similar protocols have been applied to observe spin-dependent single-atom transport in an optical lattice in real space~\cite{Karski2011}, as well as for interferometry in spin-dependent lattices~\cite{Steffen2012}.

Alternatively, the phenomena can be interpreted as helical Floquet channels, derive from windings of the Floquet quasienergy bands. More explicitly, dynamics governed by $U_h$ possess an emergent spin-rotation symmetry, preserving $S_z=(|m,a\rangle\langle m,a|-|m,b\rangle\langle m,b|)/2$.
In a perfect helical Floquet channel, pseudo-spins with spin-down (spin-up) polarization along the $z$ axis should propagate to the left (right), demonstrating a ``spin-momentum locking'' behavior discussed in Ref.~\cite{Budich2017}.
Such a behavior is clearly observed in Fig.~\ref{fig:fig4}(a)(b), where atoms initialized in $|m=0,b\rangle$ ($|m=0,a\rangle$) propagate to the left (right) in the quantum-walk dynamics.

These helical Floquet channels originate from the winding of the Floquet quasienergy bands, which is clear from the quasienergy of $U_h$ [see Fig.~\ref{fig:fig4}(c)]. Whereas the quasienergy spectrum is gapless, thus invalidating the calculation of the conventional winding numbers, the spectrum clearly shows the winding of Floquet quasienergy bands~\cite{Takuya2010}. Importantly, the decoupled linear dispersions of atoms in states $|a\rangle$ and $|b\rangle$
underlie the dispersionless helical Floquet channels observed in Fig.~\ref{fig:fig4}(a)(b), which are unattainable in static systems.
Formally, the Fourier components of $U_h$ can be decomposed into $U_{\tilde{k}}^{(\mu)}$ ($\mu=a,b$), which are Floquet operators associated with the two sublattice sites in the $\tilde{k}$ space.
Here $\tilde{k}$ belongs to the first Brillouin zone (1BZ) of the effective Hamiltonian Eq.~(\ref{eq:eff}).
The Floquet winding numbers are then defined as~\cite{Takuya2010,Budich2017}
\begin{equation}
\nu^{(\mu)}=\frac{1}{2\pi i}\int_{{\rm 1BZ}} d\tilde{k}{\rm Tr}[U^{(\mu)}_{\tilde{k}}
\partial\tilde{k}U^{(\mu)\dag}_{\tilde{k}}],\label{eq:ftpwinding}
\end{equation}
with $\nu^{(a)}=1$ and $\nu^{(b)}=-1$ under $U_h$. $\nu^{(\mu)}$ characterizes the winding of quasi-energy bands as $\tilde{k}$ traverses the 1BZ~\cite{Takuya2010}.
These Floquet winding numbers are equivalent to dynamic Chern numbers defined on the $\tilde{k}$-$t$ manifold, which are visualized as dynamic-skyrmion structures in the pseudo-spin micro-motion of the atoms, as shown in Figs.~\ref{fig:fig4}(d)(e).
Here the pseudo-spin textures are given by $\bm{n}(\tilde{k},t)=\sum_{\beta=x,y,z}\langle S_\beta \rangle_t \bm{e}_\beta$ in each $\tilde{k}$ sector ($\bm{e}_\beta$ being the unit vector of the corresponding direction).

Whereas the definition in Eq.~(\ref{eq:ftpwinding}), and hence the pseudo-spin textures in the $\tilde{k}$-$t$ space, depend on the fine-tuning of parameters, a remarkable feature of the helical Floquet channels is their robustness against perturbations.
In our experiment, the fact that we observe these channels at all relies upon this very robustness. Specifically, due to experimental error, atoms in different helical Floquet channels are inevitably coupled together. As is clear in Fig.~\ref{fig:fig4}(f), whereas the peak values of $\langle S_z\rangle_{t}$ decay below $0.1$, indicating atom loss from the corresponding helical Floquet channel, the highest peak is still on the left- (right-) most unit cell for atoms initialized in $|b\rangle$ ($|a\rangle$), such that the unidirectional flows are still observable. {We note that atom leakage from the helical Floquet channels in the presence of experimental imperfections has previously been reported for spin-dependent transport of single atoms in a spatial optical lattice.}

In order to further explore the robustness of the system against perturbations, we introduce random noise (pulse area error) in the dynamics by modulating both $\theta_1$ and $\theta_2$. In each step, $(\theta_1, \theta_2)$ pick random numbers between $(\pi/2-\Delta\theta,\pi/2+\Delta\theta)$, Here $\Delta \theta$ represents the magnitude of the random noise. At the end of 4 steps, the distribution of $\langle S_z\rangle$ is recorded.  Figure~\ref{fig:fig5}(a) and (b) show results from numerical simulations under the corresponding quantum-walk dynamics. It is clear in the figure that, with increasing noise, the unidirectional atom flow is still visible, hence robust, even for $\Delta\theta=\pi/4$, whereas the peak value in $\langle S_z\rangle$ steadily decreases. Figure~\ref{fig:fig5}(c) shows the experimental result with and without $10\%$ noise ($\Delta\theta=\pi/2\times10\%$).  It clearly shows that the distribution os $\langle S_z\rangle$ is almost the same which means the helical channel system is quite immune to noise.

\section{Conclusion}\label{sec5}

To conclude, we present a detailed analysis of the momentum lattice system with ultracold atoms. By experimentally implementing a periodically driven momentum lattice, a discrete-time quantum walk is realized, with which we observe robust helical Floquet channels. Such channels are robust to perturbations due to the topological origin of the winding of the Floquet quasi-energy bands. The periodic switching adopted here can be extended to much more complicated Floquet engineering schemes. Our experiment suggests that, the combination of momentum lattice and Floquet engineering can lead to interesting scenarios for the simulation of quantum topological phenomena. For instance, it would be intriguing to explore the interplay of Floquet topology with disorder in our many-body setting with tunable interactions.

\section{Acknowlegement}

We are grateful to Ying Hu for helpful comments. We acknowledge the support from the National Key R$\&$D Program of China under Grant No.2018YFA0307200, National Natural Science Foundation of China under Grant No. 91636104 and No. 11974331, Natural Science Foundation of Zhejiang province under Grant No. LZ18A040001, and the Fundamental Research Funds for the Central Universities. W. Y. acknowledges support from the National Key Research and Development Program of China (Grant Nos. 2016YFA0301700 and 2017YFA0304100).

\section{Author contribution}

B.Y. supervised the project. All authors contribute to the experimental data acquizition, theoretical analysis and preparation of the manuscript. 

\bibliographystyle{epj}
\bibliography{helical}

\end{document}